\documentclass[]{raa}            

\usepackage{graphicx,times}             
\usepackage{natbib}
\usepackage{amssymb,amsmath}
\bibpunct{(}{)}{;}{a}{}{,}
\usepackage[a4paper=true,dvipdfm=true,pagebackref=true]{hyperref}
\hypersetup{colorlinks = true, linkcolor = green, anchorcolor = red, citecolor = blue, filecolor = red, pagecolor = red, urlcolor = red}

\graphicspath{{img/}}
\usepackage{color}
\usepackage{ulem}

\usepackage[labelsep=space]{caption}
\usepackage{fancyhdr}

\fancypagestyle{cover}{%
	
	\lhead{}
	\rhead{Apr. 9, 2018}
	\cfoot{}}
\usepackage{upgreek}
\begin{document}
\title{Space weathering of the Moon from in situ detection
}

   \volnopage{Vol.0 (20xx) No.0, 000--000}      
   \setcounter{page}{1}          

   \author{Yunzhao Wu
      \inst{1,2,3}
   \and Zhenchao Wang
      \inst{4}
    \and Yu Lu
      \inst{5,6}
   }

   \institute{Key Laboratory of Planetary Sciences, Purple Mountain Observatory,
Chinese Academy of Sciences, Nanjing 210034, China;
   	\and
   	Space Science Institute, Macau University of Science and Technology,
Macau, China;
   	\and
   	CAS Center for Excellence in Comparative Planetology, China;
   	\and
   	Institute of Surficial Geochemistry, Department of Earth Sciences,
Nanjing University, Nanjing 210023, China;
   	\and
   	School of Geographic and Oceanographic Sciences, Nanjing University,
Nanjing, 210023, China;
   	\and
   	Jiangsu Center for Collaborative Innovation in Geographical
Information Resource Development and Application, Nanjing 210023, China;\\
\vs\no
   {\small Received~~20xx month day; accepted~~20xx~~month day}\vspace{-5mm}}

\abstract{Space weathering is an important surface process occurring on the Moon and
other airless bodies, especially those that have no magnetic field. The
optical effects of the Moon's space weathering have been largely
investigated in the laboratory for lunar samples and lunar analogues.
However, duplication of the pristine regolith on Earth is not possible. Here
we report the space weathering from the unique perspective of the
Chang$^{\prime }$E-3$^{\prime }$s (CE-3) ``Yutu'' rover, building on our
previous work (Wang et al. 2017; Wu and Hapke 2018). Measurement of the visually undisturbed
uppermost regolith as well as locations that have been affected by rocket
exhaust from the spacecraft by the Visible-Near Infrared Spectrometer (VNIS)
revealed that the returned samples bring a biased information about the
pristine lunar regolith. The uppermost surficial regolith is much more
weathered than the regolith immediately below, and the finest fraction is
rich in space weathered products. These materials are very dark and
attenuated throughout the visible and near-infrared (VNIR) wavelengths,
hence reduce the reflectance and mask the absorption features. The effects
on the spectral slope caused by space weathering are wavelength-dependent:
the visible and near-infrared continuum slope (VNCS) increases while the
visible slope (VS) decreases. In the visible wavelength, the optical effects
of space weathering and TiO$_{2}$ are identical: both reduce albedo and blue
the spectra. This suggests that developing new TiO$_{2}$ abundance algorithm
is needed. Optical maturity indices are composition related and hence only
locally meaningful. Since optical remote sensing can only sense the
uppermost few microns of regolith and since this surface tends to be very
weathered, the interpretation of surface composition using optical remote
sensing data needs to be carefully evaluated. Sampling the uppermost surface
is suggested. \vspace{-3mm}
\keywords{Space weathering; Reflectance spectra; Chang$^{\prime }$E-3
mission; In situ detection; Spectral slope}
}

   \authorrunning{Y.-Z. Wu, Z.-C. Wang, Y. Lu}            
   \titlerunning{Space weathering of the Moon from in situ detection}  

   \maketitle

%
\section{Introduction}           
Space weathering is the continuous and primary surface process occurring on
the Moon and other airless bodies, especially those that are not protected
by a magnetic field. This process is caused by various external forces,
including agglutination and disaggregation by meteoroid impacts,
implantation of ions from the solar wind, and sputtering by solar and
galactic ions and high energy photons. All of these processes cause the
regolith to mature, the degree to which is described as maturity. The
products of space weathering are agglutinitic glasses, submicroscopic iron
(SMFe), amorphous coatings on grains, new minerals, and accumulation of
meteoritic materials (e.g., Morris 1976; Keller and McKay 1997; Anand et al.
2004). The specific ferromagnetic resonance (FMR) intensity normalized to
the total iron content ($I_{s}$/\textit{FeO}) suggested by Cirlin et al. (1974), Housley
et al. (1975), Pearce et al. (1974), and Morris (1976) is the standard
measure of regolith maturity based upon the reduction of FeO to metallic Fe.
Other indices of maturity include abundance of agglutinates, concentration
of solar-wind-implanted gases, and particle size (Jolliff et al. 2006). All
these indices of maturity were derived from the investigation for bulk
soils.

Space weathering is of considerable interest because it causes changes in
the optical properties of surfaces. On the Moon, it reduces the visible and
near-infrared (VNIR) albedo and the strength of absorption bands (band depth
and band area), and increases spectral slope (Hapke 2001; Lucey et al. 2006;
Pieters and Noble 2016). Based on these optical effects, several optical
maturity indices have been created to investigate the relative age of
craters, physical processes on the lunar surface, and the reduction of
elements and minerals. Among these, the optical maturity parameter (OMAT)
proposed by Lucey et al. (2000) is the common one for remote mapping of
maturity of the lunar surface. It is the Euclidian distance from
hypothetically fully matured (dark red) origin to the point of interest in a
scatterplot of 750 nm reflectance versus 950/750 nm reflectance ratio. Other
optical maturity indices are NIR/VIS ratio (e.g., Fischer and Pieters 1996)
and the normalized visible and near-infrared continuum slope (VNCS-N) (e.g.,
Fischer and Pieters 1994; Hiroi et al. 1997; Le Mou\'{e}lic et al. 2000).

Understanding the effects of space weathering on reflectance spectra is
crucial for the interpretation of elements and minerals using remote sensing
data. Numerous experimental approaches on lunar samples (Keller and McKay
1993; Noble et al. 2001; Keller et al. 2016) and analogs (Noble et al. 2007)
have been performed to investigate space weathering. However, none of these
can realistically duplicate the actual conditions on the lunar surface. The
\textit{in situ }spectra measured by the Visible-Near Infrared Spectrometer (VNIS) onboard
the Chang$^{\prime }$E-3 (CE-3) ``Yutu'' rover provide an unique opportunity
of investigating space weathering by measuring the regolith in its
undisturbed state, as well as comparison to the regolith naturally disturbed
by rocket exhaust from the spacecraft. For example, using the CE-3 \textit{in situ} spectra,
Wang et al. (2017) estimated the SMFe abundance of the CE-3 landing site to
be 0.368 wt {\%} by means of Hapke's model. In this paper we report on
further analysis on the space weathering as measured by the CE-3 \textit{in situ }spectra.

\section{Data and Methods}

\subsection{Instrument and Data}

\begin{figure}[!htb]\centering
	\includegraphics[width=0.7\linewidth]{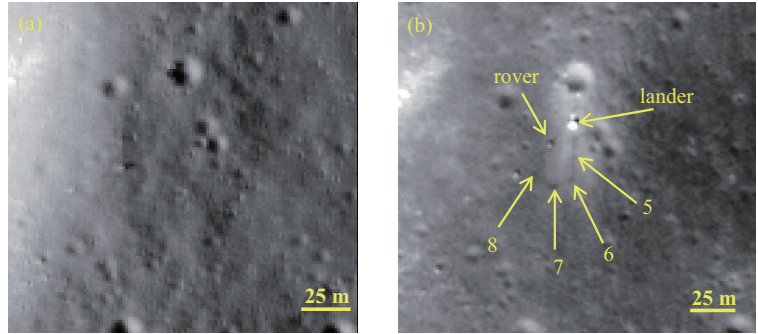}
	\caption{The Lunar Reconnaissance Orbiter Camera (LROC) Narrow Angle
Camera (NAC) images of the CE-3 landing site. (a) before landing (NAC image
M1127248516R). (b) after landing (NAC image M1147290066R).}
	\label{fig1}
\end{figure}

The VNIS consists of a VIS/NIR imaging spectrometer (450-950 nm), a
shortwave infrared (SWIR) spectrometer (900-2395 nm) and a white calibration
panel. There are 100 channels for the VIS spectrometer and 300 channels for
the SWIR spectrometer with the same spectral sampling interval of 5 nm. CE-3
was landed on the rim of Zi Wei which is in the northern Mare Imbrium. The
landing site belongs to the unsampled, spectrally unique late-stage basalts
according to their relatively strong 1000 nm feature and weaker 2000 nm
absorption compared with older basalts (Staid et al. 2011; Wu et al. 2018a).
These young basalts have high scientific significance because they provide
key information on the late-stage evolution of the Moon. Four measurements
(sites 5, 6, 7 and 8) of the soil were made by the VNIS during the period
that Yutu was mobile. The locations of the four measurements are shown in
Fig.~\ref{fig1}. Details of the instrument, data processing and calibration are
provided in Wu and Hapke (2018) and Wu et al. (2018b).

During the two lunar-day measurements by the VNIS between December 23rd,
2013 and January 14th, 2014, the farthest target measured is about 40 m from
the lander and only regoliths with no rocks were measured. For comparison,
the Indian Chandrayaan-1's Moon Mineralogy Mapper (M$^{3})$ data
(https://pds-imaging.jpl.nasa.gov/data/m3/) were also used in this study.
The M$^{3}$ is a hyperspectral instrument with 85 bands in its global mode.
In this paper, we used the Level 2 (L2) reflectance products acquired during
the Optical Period 1B (OP1B), which have the highest spatial resolution (140
m/p). The L2 data have been photometrically corrected to the standard
geometry (30$^{\circ}$, 0$^{\circ}$, 30$^{\circ})$ and corrected for thermal
emissions. Three M$^{3 }$ spectra from the landing site, Zi Wei crater, and a
350 m very fresh crater located 4.7 km southwest of the CE-3 landing site
(see Fig.~\ref{fig1}b in Wu and Hapke (2018) for detailed location) were extracted.
They represented three mature states from soils to rocks.

\subsection{Spectral Parameter Analysis}
\label{subsec:mylabel3}
Several spectral parameters including band center (BC), band depth (BD),
band area (BA), visible and near-infrared continuum slope (VNCS), visible
slope (VS), band ratio (BR) and OMAT were calculated. The VNCS is defined as
the slope of a line covering the 1000 nm absorption that connects two local
spectral maxima of its two sides. It was calculated for two types of data
and given two names for convenience: 1) the original reflectance (named
VNCS), and 2) the reflectance normalized at 750 nm (named VNCS-N). The
latter is similar to the continuum slope used previously (Fischer and
Pieters 1994; Hiroi et al. 1997; Le Mou\'{e}lic et al. 2000), which is
defined as the slope of a straight line fit tangent to both sides of the
1000 nm band and then scaled to the short-wavelength point of tangency. The
BR is 450/620 \textit{nm} for VNIS (the VS will decrease at the wavelength longer than
620 nm) and 540/750 \textit{nm }for M$^{3 }$ (the shortest band of M$^{3}$ is 540 \textit{nm}). The
OMAT was calculated using the formula in Lucey et al. (2000). Before the
spectral calculation, all the VNIS and M$^{3}$ spectra were smoothed by B
spline fitting.

To calculate BC, BD, BA and VNCS, it is necessary to derive the spectral
continuum. In this study, the straight-line continuum was used. The spectral
subset (i.e., the two endpoints) for the automatic algorithm deriving the
two tangents was assigned using the method developed in Wu et al. (2018a)
because the reflectance continuously increases in the infrared bands due to
the thermal emission from the lunar surface. The BC was calculated by
fitting a 6th order polynomial to the bottom quarter of the
continuum-removed absorption feature, and the minimum point on the
polynomial fit is considered as the BC. The BD was calculated as follows:
$D=$1-$R_{b}$/$R_{c}$, where $R_{b} $ is the reflectance at the BC and $R_{c} $ is
the reflectance of the continuum at the same wavelength as $R_{b}$. The BA
was defined as the area of the 1000 nm absorption feature for the continuum
removed spectra. The spectral slope (both VNCS and VS) was calculated as
follows: $\Delta $ reflectance/$\Delta \lambda =$
(R$_{2}$-R$_{1})$/($\lambda_{2}$-$\lambda_{1})$, where R is
reflectance and $\lambda $ is wavelength. For the VNCS, $\lambda_{2}$ and
$\lambda_{1 }$ are the wavelengths of the two tangents covering the 1000
nm absorption. For the visible slope (named VS1), $\lambda_{2}$ and
$\lambda_{1 }$ are 620 and 450 \textit{nm }for VNIS and 750 and 540 \textit{nm }for M$^{3}$.
Moreover, to avoid noise in the representation of slope with only two
endpoints, the VS was also calculated using a linear fitting between 450 and
620 nm for VNIS and 540 and 730 nm for M$^{3}$. Similar to VNCS, it was also
calculated for two types of data: 1) the original reflectance (named VS2),
and 2) the reflectance normalized at the starting wavelength (named VS-N).

\section{Results}
\label{subsec:mylabel4}
\begin{figure}[!htb]\centering
	\includegraphics[width=0.8\linewidth]{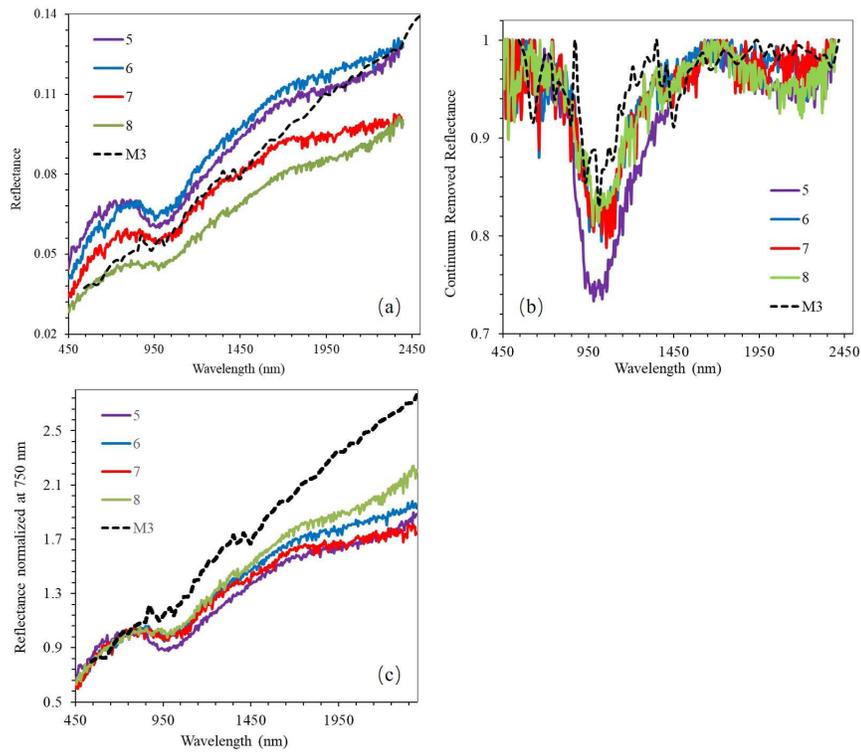}
	\caption{The VNIS reflectance spectra (a), the continuum removed
spectra (b) and reflectance spectra normalized at 750 nm (c) of 4 sites.
Also shown are M$^{3}$ spectra of regolith.}
	\label{fig2}
\end{figure}

The 4 \textit{in situ} reflectance spectra are shown in Fig.~\ref{fig2} and spectral parameters are
shown in Table 1. The \textit{in situ} spectra from all 4 sites exhibit characteristics of
lunar soil in terms of increasing reflectance with increasing infrared
wavelength and two absorptions near 1000 and 2000 nm, where the 1000 nm
absorption is relatively strong and the 2000 nm absorption is very weak. The
VNCS of the four measurements is similar. The VNCS-N, BR and BC
approximately decrease for sites closer to the lander. The reflectance,
absorption strength (BD and BA), VS, and OMAT (OMAT increases with
increasing immaturity) all increase for sites closer to the lander. The
variations of these spectral parameters of CE-3 \textit{in situ} spectra is consistent with
those of M$^{3}$ data from regolith to fresh crater.

\begin{table}
\tabcolsep=5pt
\begin{center}
\caption[]{Spectral parameters and optical maturity values of the VNIS and M$^{3 }$data. The units for all spectral slope are $\upmu m^{-1}$
and for BC is nm.}

 \begin{tabular}{cccccccccccc}
  \hline\noalign{\smallskip}
   & Site & BC & BR & BD & BA & OMAT & VNCS & VNCS-N & VS1 & VS2 & VS-N \\
\hline\noalign{\smallskip}
VNIS & 5 & 995.36 & 0.67 & 0.24 & 66.24 & 0.34 & 0.032 & 0.46 & 0.13 & 0.11 & 2.54 \\
  & 6 & 1013.50 & 0.68 & 0.17 & 44.87 & 0.24 & 0.043 & 0.64 & 0.11 & 0.12 & 3.00 \\
  & 7 & 1025.03 & 0.67 & 0.16 & 43.40 & 0.26 & 0.034 & 0.60 & 0.10 & 0.11 & 3.12 \\
  & 8 & 1020.22 & 0.72 & 0.15 & 39.91 & 0.18 & 0.033 & 0.74 & 0.07 & 0.06 & 2.25 \\
\hline\noalign{\smallskip}
M$^{3}$ & fresh crater & 994.53 & 0.76 & 0.25 & 70.98 & 0.36 & 0.014 & 0.23 & 0.07 & 0.08 & 1.73 \\
  & Zi Wei & 987.13 & 0.74 & 0.17 & 38.85 & 0.21 & 0.035 & 0.68 & 0.06 & 0.06 & 1.67 \\
  & regolith & 1010.60 & 0.78 & 0.12 & 28.67 & 0.07 & 0.052 & 1.08 & 0.05 & 0.05 & 1.49 \\
\noalign{\smallskip}\hline
\end{tabular}
\end{center}
\label{tab1}
\end{table}

\begin{figure}[!htb]\centering
	\includegraphics[width=0.5\linewidth]{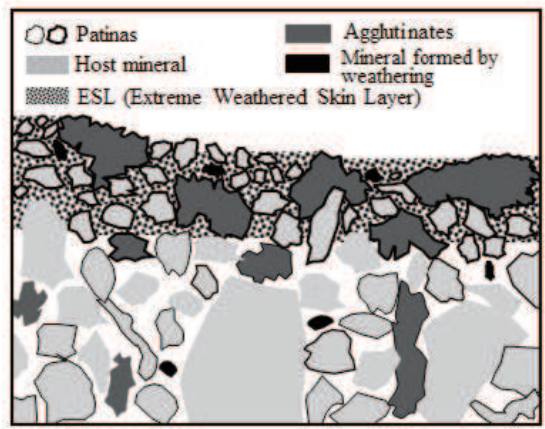}
	\caption{Model on space weathering of the real lunar surface.}
	\label{fig3}
\end{figure}

These variations of \textit{in situ} spectra between different sites indicate that the
maturity of regolith becomes less for sites closer to the lander. We propose
that this is because rocket exhaust blew away the uppermost mature dust and
soils and exposed less-mature materials. Considering that the disturbance
depth was very shallow and only the finest fraction of the regolith was
disturbed, it suggests a different model of space weathering from that
derived from lunar samples. The space weathering model from the lunar
samples shows that 1) the finest fraction of lunar soils is enriched in
Al$_{2}$O$_{3}$ and depleted in FeO and the brightest among all fractions of
lunar soils (Pieters et al. 1993; Taylor et al. 2001), 2) maturity does not
change significantly within the first tens of centimeters of regolith depth
(Clegg et al. 2014), and 3) there appears to be little difference in soil
properties inside and outside the blast zones (Clegg et al. 2014). Figure~\ref{fig3}
shows the enhanced space weathering model of lunar surface based on the CE-3
observations:

1) the uppermost surficial regolith, perhaps several millimeters to tens of
centimeters, is much more weathered than the regolith immediately below
(named as Extreme Weathered Skin Layer (EWSL) in this paper), and

2) the finest fraction is much more mature than the coarser fraction. The
weathered products are very dark and exhibit attenuated spectral features
are at all wavelengths, which mask the absorption features and reduce the
overall reflectance of covered original minerals.

The remaining materials at Site 5, that is, material not removed by the
lander rocket, have clear mafic absorptions, yet still these spectral
features are attenuated relative to fresh mineral spectra. The Site 5 \textit{in situ}
spectra also has a red VNCS. It is clear that Site 5 material contains a
significant amount of space weathered products. The BD and OMAT value of
Site 5 are similar to those of the fresh crater, and larger than those of
the wall of Zi Wei crater derived from M$^{3}$ data. The wall of Zi Wei,
which has numerous rocks, shows a similar BD but a little smaller BA and
OMAT value to Site 6. Site 8 soil has a larger BD, BA and OMAT value than
M$^{3}$ soil spectra of the landing site. These comparisons suggest that the
\textit{in situ} spectra are relatively optically immature compared to the orbital data
(i.e., the lithic fragments/mineral fraction in the pixel of \textit{in situ} spectra are
more than that of the M$^{3}$ pixel.). Further research is needed to resolve
this issue.

\section{Discussion}

\subsection{Unrepresentativeness of lunar samples for real lunar surface}

Figure~\ref{fig1} shows that the brightness of the CE-3 landing site increased after
the spacecraft landed. The increase in the reflectance of the disturbed
regolith was found for all the landing sites (Clegg et al. 2014). Smoothing
of surface roughness has been suggested as the main cause of the observed
increase in reflectance after a spacecraft has landed (Kaydash et al. 2011;
Shkuratov et al. 2013; Clegg et al. 2014). Exposure of less mature soil was
rejected in these studies based on 1) the Apollo core samples show that
maturity does not change significantly within the first tens of centimeters
of regolith depth, and 2) samples obtained using surface scoops appear to
have little difference inside and outside the blast zones at the Apollo
landing sites. The finding by VNIS that the maturity decreases immediately
below the surface is contrary to lunar sample data. Another difference is
the finest fraction. Almost all the sample spectra show that the $<$ 10 $\upmu
$m fractions are the brightest among all size fractions (Pieters et al.
1993). Sample analyses also show that for the finest fraction of the lunar
soil ($<$ 10 $\upmu $m) the abundance of plagioclase mineral fragments
increases, and is depleted in FeO and enriched in Al$_{2}$O$_{3}$ relative
to bulk lunar soil (Pieters et al. 1993; Taylor et al. 2001) that would
increase the albedo. However, as shown above the finest fractions that were
blown away are very dark and attenuated for whole wavelength, though these
finest fractions which were blown away include not only $<$ 10 $\upmu $m
fractions but also large particles. These findings indicate that the
returned samples do not represent the uppermost and most weathered layer of
lunar regolith. A visual sense compared from the terrestrial materials is
desert varnish which darkens the rock much.

\subsection{Rapid rate of SMFe formation of the skin regolith}
Recent study found that the top two centimetres of regolith is churned very
quickly, on a timescale of 81,000 years (Speyerer et al. 2016). The large
difference between spectra of the skin regolith and immediately below
derived from the \textit{in situ }spectra indicates quick development of various space
weathering products at the lunar surface compared to the vertical overturn.
Wang et al. (2017) estimated that Site 8, the visually undisturbed soil, has
0.151 wt.{\%} SMFe higher than that of the mostly disturbed soil, Site 5. It
suggests that continual bombardment of the regolith by solar wind, high
energy cosmic rays, and interplanetary dust particles result in a rapid
development of SMFe and other space weathering products (e.g., agglutinates,
mineral formed by weathering such as hapkeite (Anand et al. 2004), other
submicroscopic metals, etc).

\subsection{Reddening or bluing?}
The \textit{in situ} spectra validate the lunar-style space weathering effects and yield new
information. Canonical opinion thought that space weathering increases the
spectral slope in the visible and near-infrared (Lucey et al. 1998; Hapke
2001; Noble et al. 2001; Gillis et al. 2003), such that the 415/750 nm ratio
of Clementine data becomes smaller with increasing maturity. The \textit{in situ} spectra
show the effects on the spectral slope caused by space weathering are
wavelength-dependent: increasing the VNCS while decreasing the VS. Table 1
demonstrates that the visible spectral slope gradually decreases from
immature to mature soils, indicating that space weathering blues the surface
rather than reddens it. It is because SMFe and other space weathering
products strongly absorb light throughout the VNIR wavelengths and hence
reduces the spectral contrast (i.e., slope) between UV and visible bands. In
contrast, fresh materials and also bright materials exhibit a large spectral
contrast in the visible bands. This is caused by the strong charge-transfer
absorptions due to the Fe$^{2+}$ - Ti$^{4+}$ or O-metal transitions in the
ultraviolet extending to the visible wavelengths. This finding that space
weathering decreases the VS is consistent with the ultraviolet observations
for the Moon (Denevi et al. 2014) and asteroids (Hendrix and Vilas 2006) and
extends to the visible bands.

\begin{figure}[!htb]\centering
	\includegraphics[width=0.7\linewidth]{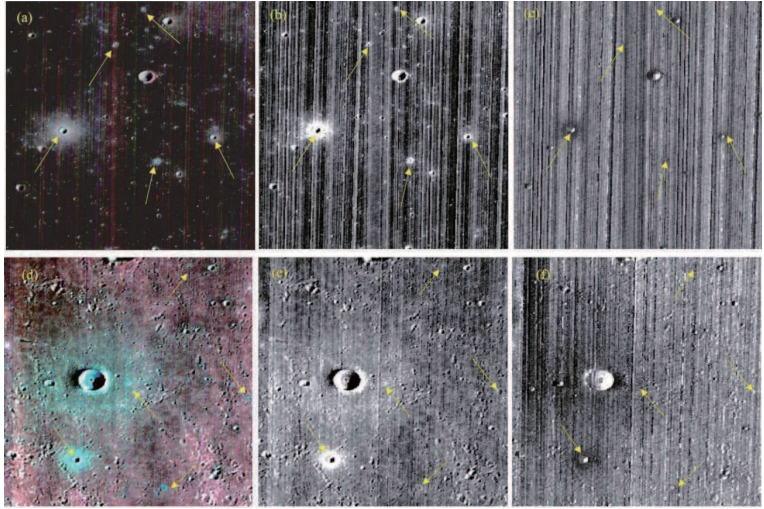}
	\caption{M$^{3}$ images showing that space weathering reduces
visible spectral slope (Fresh craters are marked with arrows). Top panel:
late-stage high-Fe basalts (similar to CE-3 landing site) with the location
of the largest fresh crater of 40.085$^{\circ}$ N, $-$26.247$^{\circ}$ W. Bottom panel: old
low-Ti basalts with the image center of 39.327$^{\circ}$ N, 0.119$^{\circ}$ E. (a) and (d)
Color composite image (R-989 nm; G-750 nm; B-540 nm). (b) and (e) Difference
image (730 nm-540 nm). (c) and (f) Ratio image (540 nm/730 nm). Note that
fresh craters show bright halo (larger slope) in the difference images (b
and e) and dark halo (lower UV/VIS ratio) in the ratio images (c and f),
indicating that fresh materials have larger visible slopes than do mature
materials.}
	\label{fig4}
\end{figure}

To validate the results from the VNIS data, we further analyzed the VS using
M$^{3}$, Clementine and LROC Wide Angle Camera (WAC) 7-channel mosaic data
for the whole Moon (supplemental material). The analysis from all three
global datasets show that immature regolith exhibits a larger VS than more
mature materials, consistent with the finding from the VNIS data. For
details, bright halos in Fig.~\ref{fig4}b {\&} e clearly indicate that the VS of
relatively young craters is larger than surrounding regolith.
Correspondingly, the 540/730 nm ratio of fresh craters is smaller than the
surrounding regolith (e.g., dark halo in Fig.~\ref{fig4}c {\&} f). Our results are in
contrast to previous results from Clementine data (Lucey et al. 1998; Gillis
et al. 2003). Since the lunar TiO$_{2}$ abundance algorithm in (Lucey et al.
1998; Gillis et al. 2003) is based on the opinion that the UV/VIS ratio
becomes smaller with increasing maturity, developing a new TiO$_{2}$
abundance algorithm is needed. Moreover, our results indicate much more
complexity in the derivation of TiO$_{2}$ regolith abundances because the
optical effects of TiO$_{2}$ and space weathering are identical, i.e.,
reducing albedo and bluing the regolith.

\subsection{Indices of Maturity}

\begin{figure}[!htb]\centering
	\includegraphics[width=0.75\linewidth]{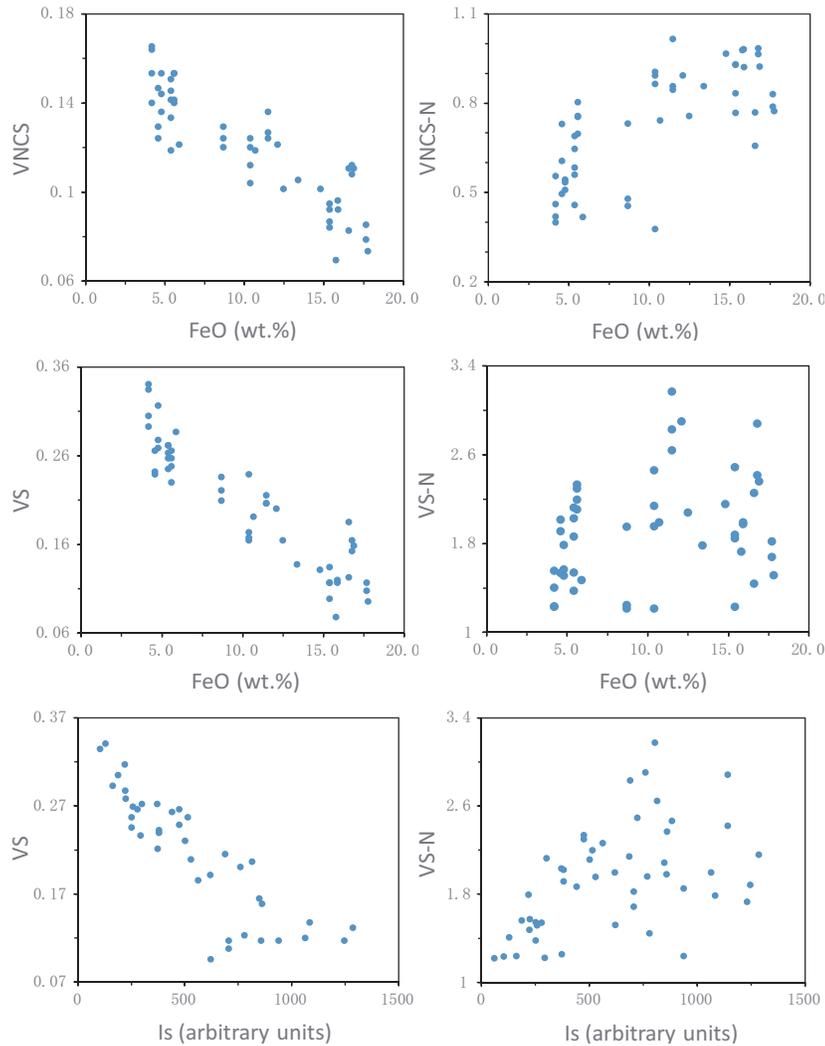}
	\caption{Scatter plots of spectral parameters and chemical compositions in lunar soils. The data are from Table 1 in Le Mou\'{e}lic et al. (2000).}
	\label{fig5}
\end{figure}

\begin{figure}[!htb]\centering
	\includegraphics[width=.8\linewidth]{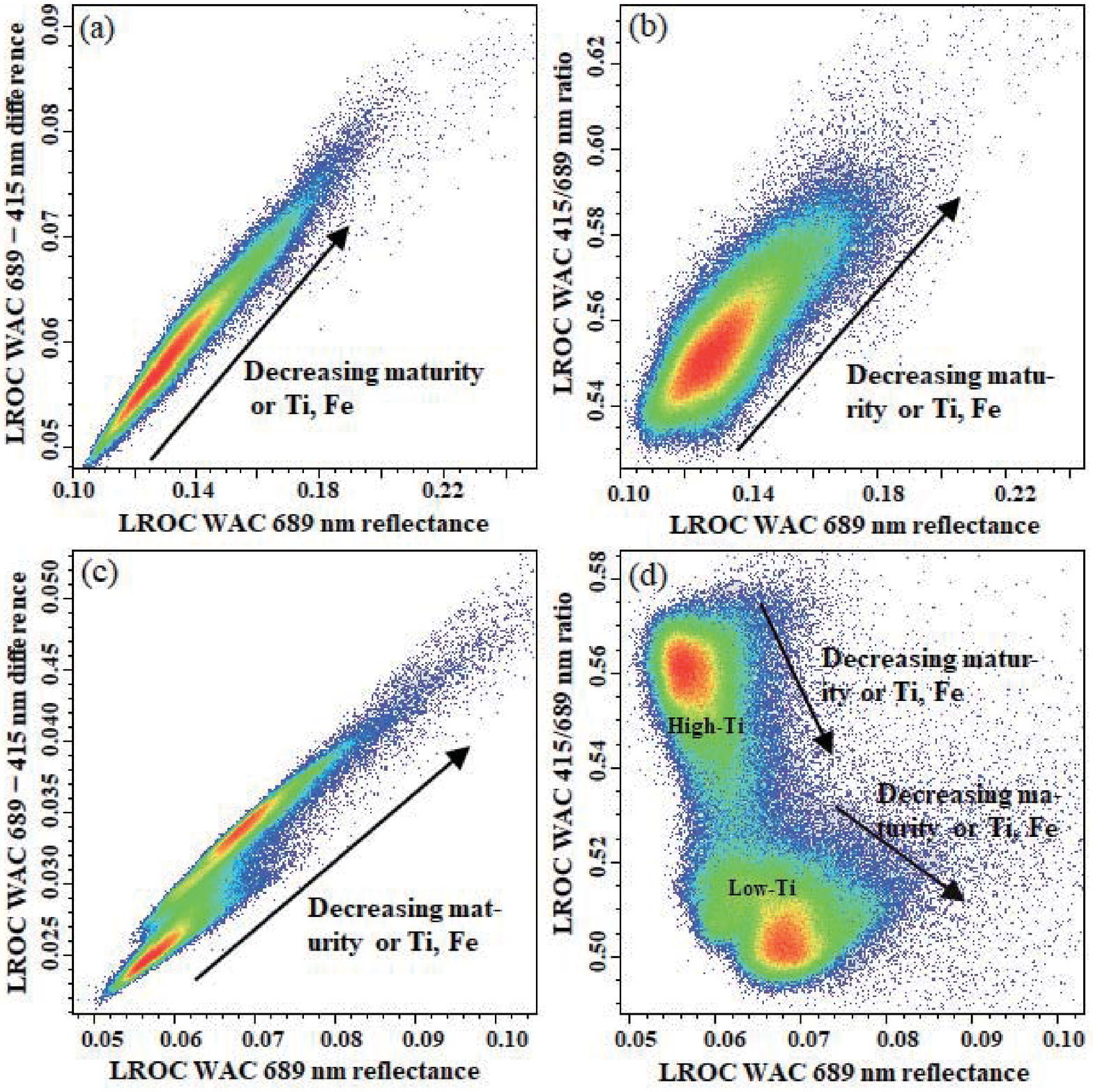}
\caption{Scatter plots of LROC WAC data for two areas (189 km * 189
km) from highlands and mare. (a) difference (689 nm -- 415 nm) vs
reflectance (689 nm) for highlands, (b) ratio (415 nm/689 nm) vs reflectance
(689 nm) for highlands. (c) and (d) are similar to (a) and (b) but for mare.}
	\label{fig6}
\end{figure}

The \textit{in situ} spectra reveal complex relationship between maturity indices (both
chemical and optical) and space weathering. Previously, spectral researches
on space weathering were focused mostly on VNCS and they concluded that the
VNCS is directly correlated with the amount of Is. Good correlations between
normalized VNCS (VNCS-N) and Is or Is/FeO have been reported in several
papers (Fischer and Pieters 1994; Hiroi et al. 1997; Le Mou\'{e}lic et al.
2000). These good correlations, however, could not be simply thought that
VNCS of lunar soils is caused by SMFe. The correlation observed by Fischer
and Pieters (1994) is for maturity index Is/FeO while that observed by Hiroi
et al. (1997) and Le Mou\'{e}lic et al. (2000) is for Is (i.e., nanophase
iron Fe) and there is no correlation with Is/FeO. The samples in Le
Mou\'{e}lic et al. (2000) show that the high FeO group has larger normalized
VNCS and Is than the low FeO group. This just indicates that VNCS is caused
by the inherent properties of the regolith (e.g., the FeO and TiO$_{2}$
abundances). For example, highlands have larger VNCS and VS than those of
maria (Fig.~\ref{fig6} in Wu et al. (2013) and Table 1 in Le Mou\'{e}lic et al.
(2000). For more detailed information see Fig.~\ref{fig5}, which shows that VNCS and
VS are negatively correlated with FeO). The continuum slope in these studies
(Fischer and Pieters 1994; Hiroi et al. 1997; Le Mou\'{e}lic et al. 2000) is
not VNCS but VNCS-N. The performance of scaling or normalization to some
band has the same influence on the optical effects as the space weathering
because the low reflectance causes a much larger slope than a high
reflectance. Figure~\ref{fig5} shows that the normalized slopes (VNCS-N and VS-N)
changed the negative correlation between FeO and slope. Table 1 and Fig.~\ref{fig2}c
also show that the VNCS of Site 8 is almost the lowest while the VNCS-N of
Site 8 become the largest because the reflectance of Site 8 is the lowest.
This resolves the discrepancy found 1) in Denevi et al. (2014) that spectral
trends related to space weathering are different for highlands and maria,
and 2) the 415/750 ratio image showing bright rays on the highlands (Fig.
S1d) and bright halos surrounding fresh craters in some areas. Figure S1d
and Denevi et al. (2014) used the reflectance ratio rather than the
difference of two bands as a measure of the spectral slope. When viewed in
terms of the difference of two bands (i.e., the mathematically true slope),
the spectral trends related to space weathering are the same for highlands
and maria, i.e., immature regolith has a larger VS than the mature
counterpart. Figure~\ref{fig6} provides more illustration. It shows that the
difference (i.e., the VS) is positively correlated with albedo for both
highlands and mare while counterpart correlations with the ratio appearing
for highlands and mare. It also illustrates that the optical effects of
TiO$_{2}$ and space weathering are identical because Ti also darkens and
blues lunar soils (e.g., Charette et al. 1974).

In addition to VNCS, OMAT has been found to correlate well with Is/FeO
(Lucey et al. 2000). However, this index also could not be simply thought
that it is related to SMFe or maturity. Firstly, as the correlation between
VNCS and Is or Is/FeO mentioned above, this correlation is not reliable
because some used Is (Le Mou\'{e}lic et al. 2000) while some used Is/FeO
(Lucey et al. 2000). The correlation plot between VNCS and OMAT (Fig.~\ref{fig4}b in
Le Mou\'{e}lic et al. 2000) show that high TiO$_{2}$ samples have smaller
OMAT values than low TiO$_{2}$ samples. This is because OMAT is more
affected by TiO$_{2}$ abundance rather than Is or Is/FeO in these data.
TiO$_{2}$ reduces spectral contrast and albedo and so reduces OMAT, which is
the same trend as the influence of space weathering. Therefore, using OMAT
to represent maturity should be used with care, taking into account the Ti
content of the surface regolith.

In summary, all the spectral parameters (e.g., brightness, VNCS, VS, BD, BA
and OMAT) describe the optical effects of space weathering and can be used
as optical maturity indices. All of them also describe the optical
characteristics of compositions. The essence of the optical effects of space
weathering is the compositional alteration. Therefore, whether for measuring
maturity or evaluating composition, the spectral parameters are only valid
locally. Previous researchers also note that OMAT works as a good measure of
relative exposure age for a regional surface but not its absolute age
because it is affected by several factors such as component and grain size
(Le Mou\'{e}lic et al. 2000; Lucey et al. 2000; Kramer 2010).

\begin{figure}[!htb]\centering
	\includegraphics[width=0.9\linewidth]{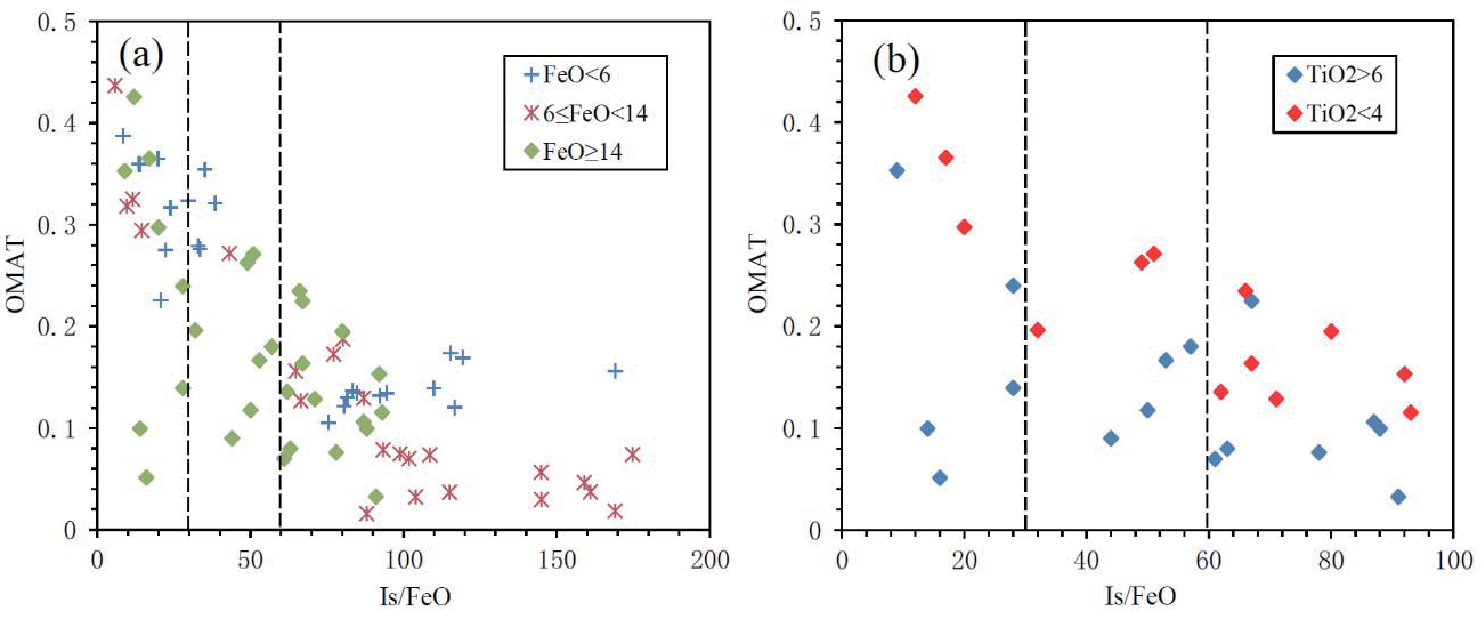}
	\caption{(a) The optical maturity parameter versus the laboratory
exposure index Is/FeO for LSCC soils. (b) the data are for samples with FeO
$>$14 wt.{\%}. The immature, submature, and mature divisions by the two
dashed lines are those of Morris (1976).}
	\label{fig7}
\end{figure}

Figure~\ref{fig7}a shows the plot of OMAT versus Is/FeO from the RELAB Lunar Soil
Characterization Consortium (LSCC)
(http://www.planetary.brown.edu/pds/LSCCsoil.html). Figure~\ref{fig7}b are the plot
for high FeO samples with FeO $>$14 wt.{\%}. As a whole, the correlation is
not apparent. Only samples with high OMAT values appear correlation and they
correspond to immature soil. For low and middle FeO samples, the OMAT
saturation respect to Is/FeO is at 70. The maximum Is/FeO for high-FeO soils
is $<$ 100, which surprisingly is much lower than low and middle FeO soils.
For high-FeO samples, the correlation is weak consistent with the FeO and
TiO$_{2}$ effects on OMAT, as discussed above. High TiO$_{2}$ soil samples
and also bright materials (see Section for the discussion of the spectral
slope) exhibit lower OMAT values than low TiO$_{2}$ soils (Fig.~\ref{fig7}b),
supporting above conclusion that TiO$_{2}$ has same optical effects as space
weathering and complicating the relationship between optical parameters and
maturity. The age of the Zi Wei crater was estimated to be $\sim $100 Ma
using its morphologic class (Basilevsky et al. 2015) or $\sim $120 Ma by
comparison of the crater morphology and rock abundances with Steno Apollo
crater (110 Ma) (Arvidson et al. 1976) and Camelot crater (143 Ma) (Drozd et
al. 1977). These analyses indicate that the CE-3 landing area is
geologically young. The OMAT of CE-3 soil, which is high FeO (22.24 wt.{\%})
and middle to high TiO$_{2 }$(4.31 wt.{\%}), is 0.18 for VNIS at Site 8 and
0.07 for M$^{3}$ data. From Fig.~\ref{fig7}b it cannot be determined whether the CE-3
soil is immature or mature. It also indicates that not only the composition
but also the data used for the calculation of OMAT (and other optical
maturity indices) can affect the optical maturity value.

\subsection{Conclusions}

The CE-3 \textit{in situ} spectra provided the unique opportunity to investigate space
weathering on the lunar surface, which could not be truly investigated in
remotely or in the laboratory. It shows that the returned samples do not
represent the pristine space weathering of the lunar regolith. The uppermost
surficial regolith, perhaps several millimeters to tens of centimeters, is
much more weathered than the regolith immediately below, indicating a rapid
development of SMFe and other space weathering products. The finest fraction
of the skin regolith is very dark and attenuated throughout the VNIR
wavelengths, indicating rich in various space weathering products. To study
space weathering of the uppermost surface in laboratory conditions would
require development of specific sampling technologies such as
electrostatically manipulating fine dust.

This study reveals that the three traditional indicators representing
maturity Is/FeO, VNCS and OMAT cannot accurately represent space weathering
of the real surface of the Moon. All the spectral parameters or optical
maturity indices are composition related and hence only valid regionally.
Moreover, the optical maturity value is also related to the data source or
instrument. The effects on the spectral slope caused by space weathering are
wavelength-dependent: the VNCS increasing while the VS decreases. At UV to
visible wavelengths, space weathering blues the regolith rather than reddens
it. It has the same sense as spectrally neutral minerals or elements (e.g.,
ilmenite or TiO$_{2})$. This suggests that the development of a new
TiO$_{2}$ abundance algorithm is needed. Since optical remote sensing can
only sense the superficial very weathered regolith, the elements and
minerals retrieved by optical remote sensing needs to be carefully
evaluated. Sampling the uppermost surface was suggested. Because the
smallest dusts are very rich in Fe$^{0}$ and easiest to be blown or
suspended, the return of these materials has implications on the
investigation of security of astronaut and rover, and contribute to the
understanding of lunar exosphere and materials transfer especially along the
terminator.

\noindent\textbf{\textit{Acknowledgements.}}
We thank the whole Chang$^{\prime}$E-3
team for making the mission a success and providing the data, Clive Neal,
Georgiana Kramer, David Blewett and Carle Pieters for discussion. We thank
two anonymous reviewers for their helpful reviews. This research was
supported by the National Key R\&D Program of China (2018YFB0504700), the Strategic Priority Research Program on Space Science, the
Chinese Academy of Sciences (XDA15020302), the Macau Science and Technology
Development Fund (103/2017/A, 119/2017/A3) and Minor Planet Foundation of
Purple Mountain Observatory.


\vspace{4cm}
\begin{center}
\textbf{Supplementary materials}
\end{center}

To confirm the VNIS results, we performed further analysis using M$^{3}$,
Clementine and LROC WAC data of nearly the whole Moon. For each dataset,
both the ratio image (reflectance of short wavelength divided by that of
long wavelength) and the difference image (reflectance of long wavelength
minus that of short wavelength) were calculated. Figure S1 shows the
results. The three datasets show that space weathering consistently reduces
VS rather than increasing it. (Note that the difference of two bands
represents the spectral slope while the ratio of two bands does not
represent the spectral slope but the spectral normalization.) Moreover, to
avoid the noise of two individual bands, an alternative method of
calculation of the spectral slope can be a linear fit using multiple bands.
This gives the same results as the difference of two bands.

\begin{figure}[!htb]\centering
	\includegraphics[width=0.6\linewidth]{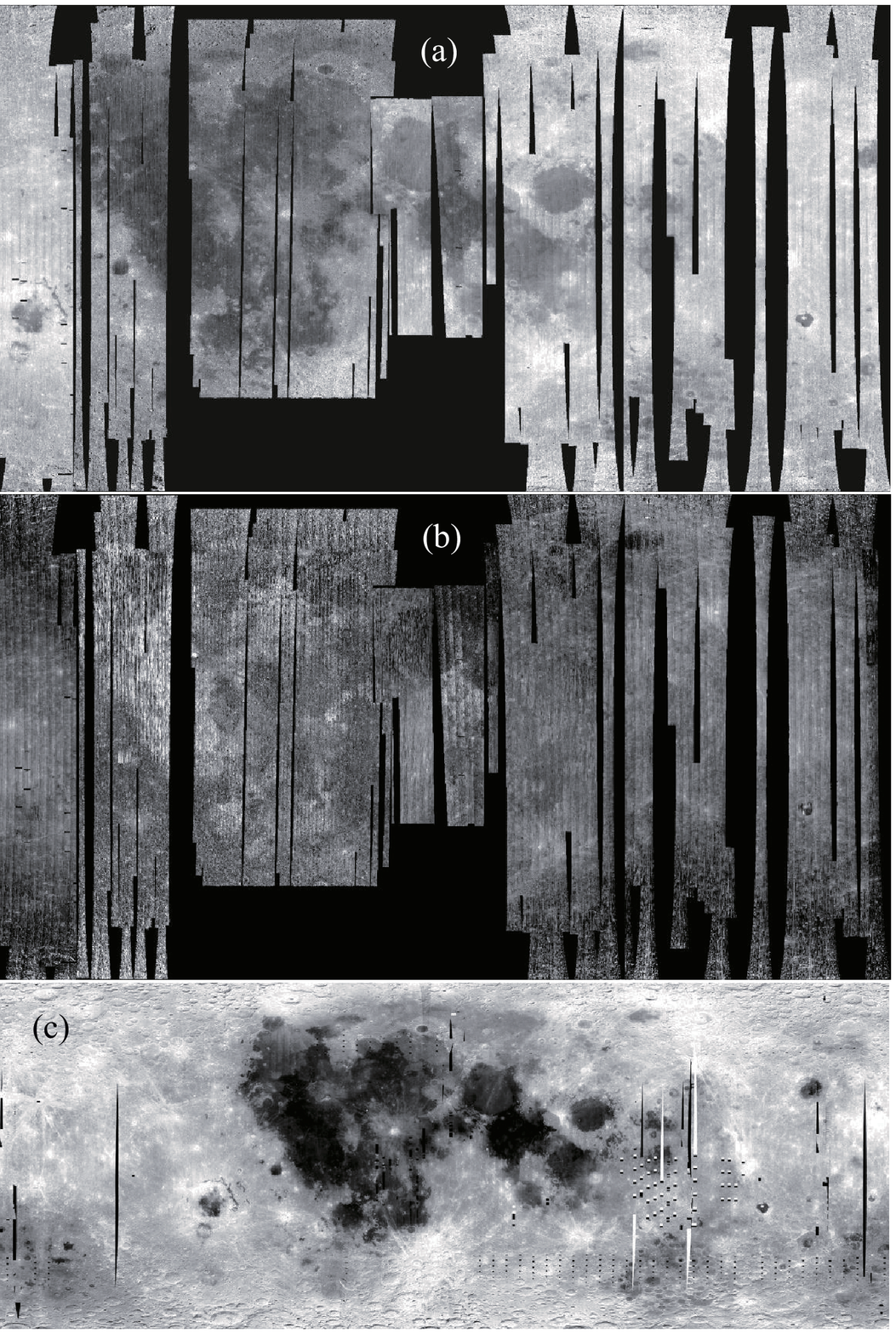}\\
\includegraphics[width=0.6\linewidth]{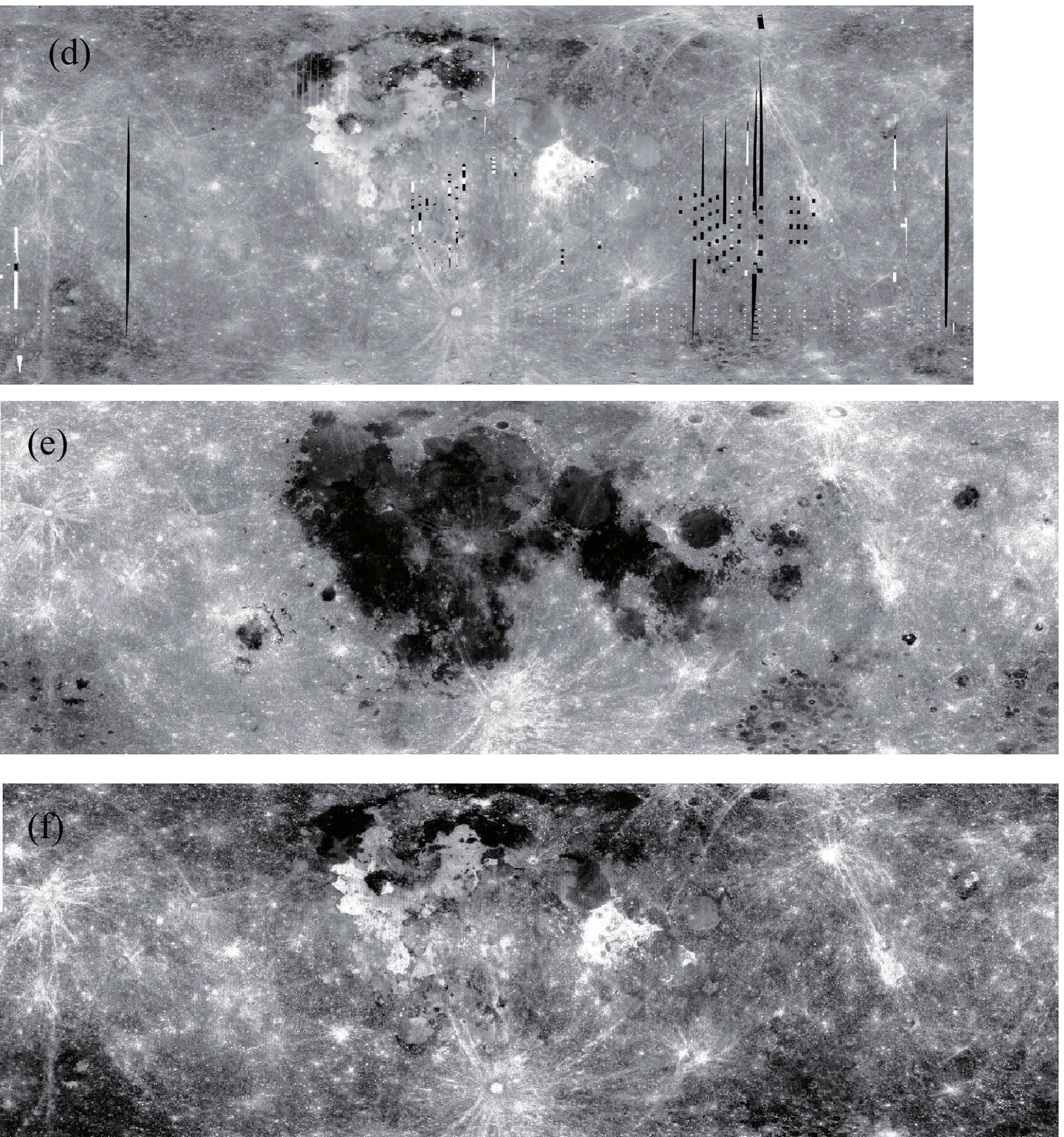}
	\caption*{Fig. S1. Difference image (730 nm -- 540 nm) (a) and ratio image
(540 nm/730 nm) (b) of global M$^{3}$ data. (c) and (d), and (e) and (f) are
similar to (a) and (b) but for Clementine (750 nm and 415 nm) and LROC WAC
(689 nm and 415 nm) data.}
	\label{figS1}
\end{figure}

\end{document}